\documentclass[twocolumn,amsmath,amssymb,nofootinbib,floatfix]{revtex4}

\usepackage{graphicx}
\usepackage{dcolumn}
\usepackage{bm}
\usepackage{longtable}
\usepackage{float}

\usepackage{subfigure}
\usepackage{array}
\usepackage{pstricks,pst-node,pst-plot}
\usepackage{pst-grad}
\usepackage{tabularx}

\newpsobject{grilla}{psgrid}{subgriddiv=1,griddots=10,gridlabels=6pt}

\usepackage{color}
\usepackage{ulem}

\newcommand{\troc}{{\footnotesize \bf T\raisebox{-0.6ex}{R}O\raisebox{-0.6ex}{C}A\raisebox{
-0.6ex}{D}E\raisebox{-0.6ex}{R}O}}
\newcommand{\be}{\begin{equation}}
\newcommand{\ee}{\end{equation}}

\newcommand{\bea}{\begin{eqnarray}}
\newcommand{\eea}{\end{eqnarray}}

\newcommand{\bt}{\begin{tabular}}
\newcommand{\et}{\end{tabular}}
\newcolumntype{P}{>{\footnotesize}c<{}}
\newcolumntype{M}{>{\scriptsize}c<{}}
\newcolumntype{N}{>{\scriptsize}l<{}}

\mathchardef\mhyphen="2D
\begin{document}

\preprint{PREPRINT}

\title{Computation of the thermal conductivity using classical and quantum  molecular dynamics based methods.}

\author{Natalia Bedoya, Jean-Louis Barrat}
 \affiliation{
   Nanosciences Foundation, 23 rue des martyrs, 38000 Grenoble, France,
   Laboratory for Interdisciplinary Physics, UMR 5588, Universit\'e Grenoble 1 and CNRS, Saint Martin d'H\`eres, 38402, France
     }
\author{David Rodney}
 \affiliation{
  SIMaP, Grenoble INP, UJF, CNRS, UMR 5266, Saint Martin d'H\`eres, F-38401
  France }

\date{July 8, 2013}

\begin{abstract}
The thermal  conductivity of a  model for solid argon  is investigated
using  nonequilibrium  molecular  dynamics  methods, as well as  the
traditional  Boltzmann  transport  equation  approach  with  input  from
molecular  dynamics  calculations,  both  with classical  and  quantum
thermostats. A  surprising result is  that, at low  temperatures, only
the classical  molecular dynamics technique  is in agreement  with the
experimental  data.  We  argue  that   this  agreement  is  due  to  a
compensation of errors,  and raise the issue of  an appropriate method
for   calculating  thermal   conductivities  at   low   (below  Debye)
temperatures.
\end{abstract}


\maketitle

\section{Introduction}
\label{Intro}

Different methods are available for calculating thermal conductivities
of  crystalline  solids.~\cite{RMG2010_Stackhouse}  The most  standard
approach  involves  a calculation  of  the  phonon  properties of  the
system, which  are connected to  the thermal conductivity  through the
Boltzmann  transport  equation.   Alternative  methods  are  based  on
equilibrium  molecular  dynamics  (MD) and  non-equilibrium  molecular
dynamics (NEMD).~\cite{PRB1999_Jund,JCP1997_Plathe,RMG2010_Stackhouse}

 The  methods based  on MD  or NEMD  are restricted  to  the classical
limit, i.e.  the  limit of high temperatures.  In  standard MD, nuclear
degrees of  freedom are treated classically and  quantum effects such
as  zero-point vibrations  are  not accounted  for.  In  order to
incorporate  quantum  effects,  corrections   to  the  thermal
conductivity, based on a rescaling  of the heat capacity, are commonly
applied.   These  kind  of  corrections, however,  are  not  generally
accepted as reliable.  Turney et al.~\cite{Turney2009a} discussed
their validity,  and  showed  that this approach  is oversimplified
and is  not generally  applicable, while other  authors have  found an
improvement  of the  classical thermal  conductivity by  applying such
corrections.~\cite{Li1998,Jund1999}  Quantum  effects,  on  the  other
hand,  are  assumed  negligible  depending  on the  capability  of  the
classical  description  to  describe  the  thermal  conductivity,  but
independently  of  its limitations  for  predicting  heat capacities  or
phonon  lifetimes,  properties   directly  related  with  the  thermal
conductivity.  A common  example is the case of  solid argon. In spite
of the  limitations of the  classical theory to predict  correctly the
heat capacity, a reliable  description of the thermal conductivity, at
temperatures well  below the Debye  value, is obtained  from classical
molecular dynamics.~\cite{kaburaki_JAP_07,McGaughey2004}

Quantum  effects on  the  thermal conductivity  can  be obtained  from
anharmonic  lattice   dynamics,  by  using   the  Boltzmann  transport
equation.~\cite{srivastava} This  methodology, nonetheless, results in
much more expensive computations than molecular dynamics.  It requires
the full  calculation of  the vibrational spectrum  of the  system, as
well as  the third derivatives  of the energy,  something unmanageable
for  large or aperiodic systems.  Moreover, the  Boltzmann transport
calculation  relies  on approximate  theoretical  expressions for  the
phonon lifetime and for the  conductivity itself, as opposed to the MD
and NEMD formalisms that are in principle exact.

 Recently, a Langevin  type thermostat with a coloured  noise was
proposed         ~\cite{PRL2009_Dammak,PRL2009_Parrinello}         and
implemented  \cite{,JSP2011_Barrat_Rodney}  by  different authors  in
order  to  incorporate quantum  effects  in  molecular dynamics.   The
quantum thermostat  allows  to recover the  correct average quantum
energy of a system by coupling every degree of freedom to a fictitious
quantum bath,  in such  a way that  a harmonic oscillator  acquires an
energy given by the Bose-Einstein  distribution. As such the method is
expected  to provide  a good  description  of solids  in the  harmonic
limit,  and has  been  shown to  also  work well  for low  temperature
liquids, in terms  of static properties.~\cite{PRL2009_Dammak} At high
temperatures, the  quantum thermostat  reduces to a  standard Langevin
thermostat.   This semi-classical approach  offers the  possibility of
performing   direct  thermal   conductivity  calculations,   by  using
molecular dynamics,  independent of the temperature  regime.  Savin et
al.~\cite{Savin2012}  have applied  this methodology  to the  study of
heat transport in  low-dimensional nanostructures from non-equilibrium
molecular dynamics (NEMD).  In the case of a NEMD simulation there are
regions of the  system free of thermostat, and one  will have to check
the  validity  of  the   quantum  thermostat  under  such  conditions.
Moreover, the quantum thermostat is not an exact representation of the
quantum behavior,  and for anharmonic  systems suffers from
``zero-point energy leakage'' (see Ref.
\onlinecite{PRL2009_Parrinello} and section \ref{thermostat} below). It is
unknown if this influences thermal  transport properties.   Here we
present an overview of the advances and challenges for using such
kind of  thermostats to address  thermal transport studies  at low
temperatures (below the Debye temperature $T_\text{D}$).  In our study
we  use  different  MD  based  methods  for  calculating  the  thermal
conductivity of solid argon, a simple system that is well described in
the  literature  and, as  pointed  out  before,  is particularly  well
described by classical MD.

The  paper is  organized as  follows.  In  Sec.~\ref{methods}  we 
present  the   various  methods   used  for  estimating   the  thermal
conductivity.   In  Sec.~\ref{thermostat}  we  briefly  introduce  the
quantum  thermostat, we  discuss about  the zero-point  energy leakage
problem  and  its  reliability   for  working  under  equilibrium  and
non-equilibrium conditions.   In the last section we  present and
discuss our results for the thermal conductivity of solid Argon.

\section{Methodology}
\label{methods}
The  standard  methods to  compute  thermal  conductivities are  based
either on molecular dynamics or  lattice dynamics, or a combination of
both.  In  this work, we  used  non-equilibrium molecular dynamics
(NEMD)~\cite{PRB1999_Jund,JCP1997_Plathe,RMG2010_Stackhouse}        and
Boltzmann       transport       equation      molecular       dynamics
(BTE-MD).~\cite{PRB09_turney,McGaughey2004b}   We   did   not   use
Green-Kubo based methods,~\cite{Green1954,Kubo1957} which have smaller
size effects than NEMD, because  the quantum thermostat is not compatible 
with this approach (see  discussion below). However, the cells employed
here are large enough to avoid any strong size effects in NEMD.

In NEMD, the periodic simulation cell is divided into $N$ slabs, and a
temperature gradient is imposed by  coupling two selected slabs to two
thermostats at  different temperatures, $T_\text{1}$  and $T_\text{2}$
with $T_\text{1}<T_\text{2}$.  In a periodic system, the thermostated
slabs are separated by a distance  equal to one half of the simulation
cell length. The remaining slabs  are not thermostated.  The system is
then  allowed to reach  a steady  state, where  on average  the energy
creation rate of the thermostat at $T_\text{2}$ is equal to the energy
removal rate of the  thermostat at $T_\text{1}$.  Calculating the heat
flux  $j_i$   required  to   maintain  the  gradient   of  temperature
${\nabla}_jT$ from the heat power of  the source and the sink, one can
estimate the thermal conductivity from Fourier's law:
\begin{equation}
j_i=-\kappa_{ij}\,{\nabla}_j\,T.
\label{eq:kappa}
\end{equation}
In this  work we assumed  materials of isotopic symmetry.  The thermal
conductivity is then  a scalar, and the temperature  gradient and heat
flux are parallel.

Equation (\ref{eq:kappa})  can alternatively be implemented  by imposing the
heat  flux   $\vec{J}$  and  calculating   the  resulting  temperature
gradient.        A        common        approach        in        this
case, \cite{PRB1999_Jund,JCP1997_Plathe} is to rescale the velocities,
$\vec{v}_\text{h}$, of the atoms in the hot region according to
\begin{equation}
\vec{v}_\text{h}\,' = \vec{v}_\text{G}+\alpha(\vec{v}_\text{h}-\vec{v}_\text{G}),
\label{eq:velocity_scale}
\end{equation}
where $\vec{v}_\text{G}$ is the velocity of the center of mass of the region, and 
 \begin{equation}
\alpha=\sqrt{1+\frac{\Delta\,\epsilon}{k_\text{R}}}.
\label{eq:alpha_jund}
\end{equation}
Here $\Delta\,\epsilon$  is the amount of heat  transferred through the
system, and $k_\text{R}$ is the relative kinetic energy given by
\begin{equation}
k_\text{R} = \frac{1}{2}\sum_{i \in \text{hot}} m_i\;\vec{v}_i^{\;2}-\frac{1}{2}\sum_{i \in \text{hot}} m_i\;\vec{v}_\text{G}^{\;2}.
\label{eq:kinetic_jund}
\end{equation}
In this manner, a constant heat flux 
\begin{equation}
J=\frac{\Delta\,\epsilon}{2\,A\,\Delta\,t}
\end{equation}
is imposed,  where $A$ is the  cross sectional area  of the simulation
cell  perpendicular to  the  heat flow,  and  $\Delta t$  is the  time
step. We implemented both NEMD methods and  checked that they
are fully consistent  with one another. In the  following, we will not
distinguish between  them and  will simply refer  to them as  the NEMD
approach.

An alternative expression for the thermal conductivity of an isotropic
material reads~\cite{srivastava,PRB09_turney}
\begin{equation}
\kappa=\sum_{\vec{q}}\sum_\nu^{3\,(N-1)} C_\text{ph}(\vec{q},\nu)\,{v}_\text{g}^{\,2}(\vec{q},\nu)\,\tau(\vec{q},\nu),
\label{eq:BT}
\end{equation}
where   $C_\text{ph}$  is   the  volumetric   phonon   specific  heat,
$\vec{v}_\text{g}$ is the phonon  group velocity and $\tau$ the phonon
lifetime.  The  sum runs over  all wave vectors, $\vec{q}$,  within the
Brillouin  zone  of  the  periodic  structure,  and  over  the  $3\,N$
polarization  indices,  where  $N$  is  the number  of  atoms  in  the
elementary cell  under consideration,  so that contributions  from all
normal modes of the system are considered.  In Eq.  (\ref{eq:BT}), the
specific  heats and  group velocities  can be  computed  using lattice
dynamics,  while the  phonon lifetimes  can be  obtained  using either
lattice dynamics  or a combination  of lattice dynamics  and molecular
dynamics.~\cite{PRB09_turney,srivastava} In the following, we will use
only the latter approach, referred to as the BTE-MD method.

Group  velocities are  obtained by  evaluating the  derivative  of the
dispersion curves,  $\omega(q)$, over a  set of 6 $q$-points  within a
radius  0.0001   $\AA^{-1}  $  around  the   $\Gamma$  point.   Phonon
frequencies and  specific heats are calculated at  the $\Gamma$ point,
in supercells that contain from 256 to 4000 atoms.

Phonon  lifetimes  are  obtained  from  the  energy  autocorrelation
function of each normal mode
\begin{eqnarray}
E^{\,\vec{q},\nu}(t)&=&\frac{Q^{\ast}(\vec{q},\nu) \, Q(\vec{q},\nu)}{2}\,+\\
&&\frac{\omega^{\,2}\,(\vec{q},\nu) \; \dot{Q}^{\ast}(\vec{q},\nu) \, \dot{Q}(\vec{q},\nu)}{2}, \nonumber
\end{eqnarray}
with 
\begin{eqnarray}
 Q(\,\vec{q},\nu)&=&\sum_{j=1}^N \,\left[\frac{m_j}{N} \right]^{1/2}\text{exp}[{-i\vec{q}\cdot\vec{r}_{j,0}}] \, \times\\ && \,\vec{e}\,(\vec{q},\nu)\cdot[\vec{r}_{j}-\vec{r}_{j,0}]\nonumber
\end{eqnarray}
the   time-dependent   normal   mode  coordinate.   The   eigenvectors
$\vec{e}({\,\vec{q},\nu})$ are  obtained from lattice dynamics, and
the relative  displacement, $\vec{r}_{j}-\vec{r}_{j,0}$, of  atom $j$,
is sampled using molecular dynamics. The phonon lifetimes $\tau$ are then
obtained by fitting the following relation
\begin{equation}
e^{-t/\tau(\,\vec{q},\nu)}=\frac{\langle E^{\,\vec{q},\nu}(t)\, E^{\,\vec{q},\nu}(0) \rangle}{ \langle E^{\,\vec{q},\nu}(0)\, E^{\,\vec{q},\nu}(0) \rangle}.
\end{equation}

\section{Quantum thermostat}
\label{thermostat}
\subsection{Overview}
The key idea  behind the quantum thermostat  is to adjust to the
manner in  which energy  is distributed among  the normal modes  of a
harmonic system. In the  classical limit, the equipartition theorem is
fulfilled and  all modes  have the same  energy, while in  the quantum
regime,  the   energy  of  each  mode  is   distributed  according  to
Bose-Einstein  statistics. The  quantum  Langevin thermostat  enforces
this  distribution  by  using  a  frequency-dependent  noise  function
(coloured noise).

As  in  the  classical  approach  using a  Langevin  thermostat,  each
particle is coupled to a fictitious bath by including in the equations
of  motion  a random  force  and a  dissipation  term  related by  the
fluctuation-dissipation    theorem.~\cite{risken}   Accordingly,   the
equation of motion  of a degree of freedom $x$, of  a particle of mass
$m$, in presence of an external force $F(x)$, becomes
\begin{equation}
m\,\ddot{x}=-\gamma\,m\;\dot{x}+F(x)+{\sqrt{2\,m\,\gamma}\,\Theta(t)},
\label{eq:langevin}
\end{equation}
where  $\Theta(t)$ is  a coloured  noise with a power  spectral density
(PSD) given by the Bose-Einstein distribution
\begin{eqnarray}
\label{eq:omega}
\tilde{\Theta}(\omega)&=&\int e^{-i\omega\,t} \langle \,\Theta(t) \,\Theta(t') \,\rangle d\,t\\ \nonumber
                      &=&\hbar\,|\omega|\left(\frac{1}{2}+\frac{1}{e^{\,\hbar\,|\,\omega\,|\,k_\text{B}\,T}-1}\right),
\end{eqnarray} 
including the zero point energy.  The classical regime is recovered at
high  temperature, where  the  above PSD  becomes  independent of  the
frequency and equals  $k_\text{B}\,T$.~\cite{Grest1986} We also note
here that the use of a  Langevin equation implies the absence of local
energy conservation; hence a  Green-Kubo approach, based on the notion
that  local energy  fluctuations undergo  a diffusive  motion,  is not
appropriate  in  a system  that  is coupled  to  a  local (quantum  or
classical) heat bath.

In practice,  $\tilde{\Theta}(\omega)$ is generated by  using a signal
processing     method      based     on     filtering      a     white
noise.~\cite{Oppenheim2009} A filter
\begin{equation}
\tilde{H}(\omega)=\sqrt{\tilde{\Theta}(\omega)}
\end{equation}
with Fourier transform $H(t)$ is  introduced, and  $\Theta(t)$
is  obtained by convoluting  $H(t)$ with  a random
white    noise,    $r(t)$,   of    power      spectral    density
$\tilde{R}(\omega)$=1, such that
\begin{equation}
\Theta(t)=\int\limits_{-\infty}^{\infty} H(s)\,r(t-s)\,ds.
\label{eq:convolution}
\end{equation}
Thus, the power spectral  density of the resulting  noise is
\begin{equation}
\left| \tilde{H}(\omega) \right|^2\tilde{R}(\omega)=\tilde{\Theta}(\omega),
\end{equation}
which satisfies Eq. \!(\ref{eq:omega}).   The method is simple and can
be  easily implemented  in  a discrete  molecular dynamics  algorithm.
From a  computational point of  view, the quantum thermostat  does not
slow  down the  calculations,  the only  difference  with a  classical
thermostat      being      the      convolution      operation      in
Eq.  \!(\ref{eq:convolution}).   In   terms  of  memory,  the  quantum
thermostat is  more demanding  because it requires  to store  a finite
number of past values of the white noise and of the filter in order to
compute Eq. \!(\ref{eq:convolution}).  However, the memory requirement
for the thermostat scales linearly  with the system size and is easily
manageable with current computers.  Moreover, it avoids generating and
storing  the entire  time-series of  random numbers  as done  in other
implementations  of  the  quantum thermal  bath.~\cite{PRL2009_Dammak}
Further    details    concerning    the    method   are    given    in
Ref.~\onlinecite{JSP2011_Barrat_Rodney}.

\subsection{Zero-point energy leakage}
\label{sec:zpe}
By coupling a system to the quantum thermostat,  each harmonic mode can
in principle  be equilibrated at  the correct quantum  harmonic energy
given  by Eq.  (\ref{eq:omega}). However,  as the  equations  that are
solved describe classical coordinates,  the zero point energy in these
equations corresponds to the finite amplitude vibration of a classical
coordinate. As  such this zero  point energy can be  exchanged between
modes, in contrast with a true quantum zero point energy.

\begin{figure}
\includegraphics[scale=0.4]{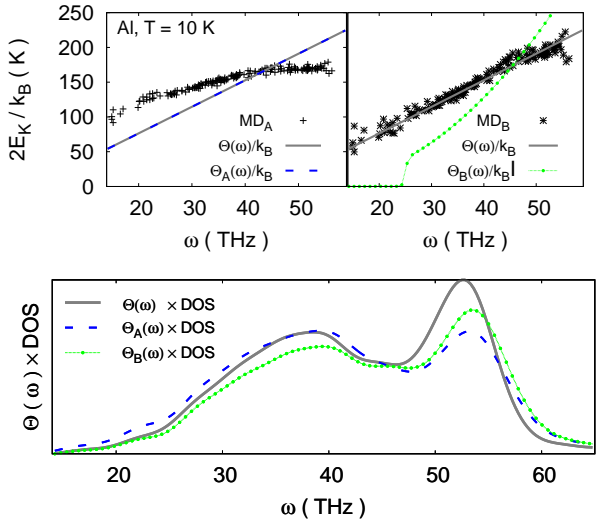}
\caption{ Scaled kinetic energy distribution per  mode obtained from MD simulations.  Left
  top  panel: values  obtained  using a  noise with  power spectral
  density   $\tilde{\Theta}_A(\omega)$  equal  to   the  Bose-Einstein
  distribution $\tilde{\Theta}(\omega)$ [Eq. (\ref{eq:omega})].  Right
  top panel: values obtained using a noise with power spectral density
  $\tilde{\Theta}_B(\omega)$.    Bottom   panel:  vibrational   phonon
  spectrum times the phonon density  of state.  The full gray line was
  obtained by calculating  the DOS using a classical  thermal bath and
  multiplying  by   $\tilde{\Theta}(\omega)$.  The  other   data  were
  obtained  directly  from  the  Fourier  transform  of  the  velocity
  auto-correlation  function (VAF).  Calculations  were performed  for
  solid aluminum at 10 K using a Lennard-Jones potential.}

\label{fig:leakage}
\end{figure}

Such an exchange becomes  possible when an anharmonic coupling between
the modes is introduced, and  leads to the phenomenon of ''zero-point
energy  leakage'' (ZPE), where  the zero  point energy  is transferred
from the  high-energy modes to  low-energy modes, so as  to homogenize
the                  energy                  among                 the
modes.~\cite{Miller1989,Alimi1992,Ben-Nun1996,Habershon2009,Czako2010}
As  the  thermostat  cannot   fully  counterbalance  the  leakage,  an
equilibrium is reached where the  energy per mode is neither constant,
nor as inhomogeneous as  in Bose-Einstein distribution.  An example is
shown in Fig.~\ref{fig:leakage} (left panel)  in the case of a perfect
crystal of aluminum at 10  K modeled with a Lennard-Jones potential. A
coloured noise with power spectral density $\tilde{\Theta}_A(\omega) =
\tilde{\Theta}(\omega)$  directly from  Eq. (\ref{eq:omega})  was used
and results in  an excess of energy in modes  with frequency less than
about 40  THz and a deficiency  in energy for modes  above that value.

One way  to correct for  the leakage is  to modify the  power spectral
density of the  filter, such that after equilibration  of the leakage,
the  system  reaches an  energy-mode  distribution  which follows  the
Bose-Einstein    distribution.     An    example    is     shown    in
Fig.~\ref{fig:leakage} (right panel)  with the adjusted power spectral
density,  $\tilde{\Theta}_B(\omega)$,  shown  as  a dotted  line.  The
resulting energy distribution (asterisks)  is in much better agreement
with  the  desired Bose-Einstein  distribution  (full  line) than  the one obtained using 
the original filter, shown  in the left panel.  The  leakage is however not
perfectly corrected, as can be  seen from the Fourier transform of the
velocity     auto-correlation      function     (VAF)     shown     in
Fig.~\ref{fig:leakage} (bottom panel).  Bear in mind that in classical
MD, i.e.   when a thermostat  fulfilling the equipartition  theorem of
energy is  used, the  Fourier transform of  the VAF equals  the phonon
density of  states (DOS) times $k_\text{B}\,T/m$; in
the quantum case, we obtain the phonon DOS times the phonon population
function  [Eq.  (\ref{eq:omega})].   Figure~\ref{fig:leakage} compares
results obtained with the two coloured noises, $\tilde{\Theta}_A(\omega)$
and $\tilde{\Theta}_B(\omega)$, to the exact distribution.  The latter
is  estimated by calculating  the vibrational  DOS, using  a classical
thermal  bath (CTB), multiplied  by $\tilde{\Theta}(\omega)$.   In the
case  where  $\tilde{\Theta}_A(\omega)=\tilde{\Theta}(\omega)$ (dashed
blue  line), the  ZPE leakage  results  in an  underpopulation of  the
high-energy  modes  ($\omega>45$ THz)  and  an  overpopulation of  the
low-energy      modes.      The      corrected     coloured      noise
$\tilde{\Theta}_B(\omega)$ (green  dotted line) yields  a much better,
although  not  perfect,  agreement  with the  exact  distribution,  but
fitting such  power spectral density is technically  difficult. The
corrected PSD  is both  system- and temperature-dependent,  making the
direct application of this procedure rather tedious.

In  spite   of  the  zero  point  energy   leakage,  quantum  Langevin
thermostats   have   been   successfully   used   to   map   out   the
diamond-graphite coexistence  curve,~\cite{PRL2009_Parrinello} as well
as    the   proton    momentum   distribution    in   hydrogen-storage
materials.~\cite{PRB2010_Parrinello_H}  The   quantum   effects
  accounted for the thermostat were relevant, in these cases,  for a
  correct  description of the  systems. Expecting  the same  degree of
  accuracy   of  the  method   for  describing   thermal  conductivity
  properties, our calculations  were performed omitting any correction
  concerning  the leakage.  However,  as will  be  shown later,  this
  introduces a serious  limitation for the method.

\subsection{Quantum thermostat and NEMD}
\label{sec:QT-NEMD}
\begin{figure}
\includegraphics[scale=0.4]{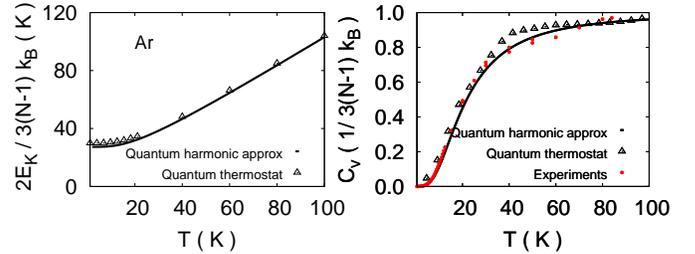}
\caption{Left panel:  Scaled average kinetic energy per  degree of freedom,
  i.e. by  definition the classical temperature, $T^\text{C}$,  as a function
  of  temperature.   Right  panel:  Specific heat  as  a function  of
  temperature   obtained   from   molecular   dynamics   compared   to
  experimental    data~\cite{cv_exp}   and   the    quantum   harmonic
  approximation. Within the harmonic approximation we have $C_\text{v}
  =              \frac{\,\,\,\,d\,T^\text{C}}{d\,T}$,              see
  Eq.~\ref{eq:temperatures}.   The phonon  spectrum was  obtained from
  lattice dynamics  calculations performed at  the $\Gamma$-point, using a
  1280-atoms cell.}
\label{fig:cv_ek}
\end{figure}

 Despite the  zero-point energy  leakage described above,  the quantum
 thermostat allows  to recover  the correct temperature  dependence of
 the  equilibrium average  thermal energy  and heat  capacity.  Figure
 \ref{fig:cv_ek} shows  an example  in the case  of solid  argon. Here
 every degree of freedom of the system is coupled to the thermostat.
 
In the present semi-classical  approach, we should distinguish between
the temperature  used as  input of the  thermostat, which is  the true
(quantum)  temperature  of  the   system,  denoted  as  $T$,  and  the
temperature measured from  the kinetic energy of the  system, which we
call the  classical temperature, $T^\text{C}$. The relation  between both, in
the case  of solid  argon, is shown  in Fig.  \ref{fig:cv_ek}.  In the
harmonic approximation, we have:
\begin{equation}
\label{eq:temperatures}
T^\text{C} = \frac{1}{3(N-1)k_\text{B}}\sum\limits_{i}^{3(N-1)}\hbar\,\omega_i \,\left(\frac{1}{2}+\frac{1}{e^{\,\hbar\,\,\omega_i\,/ \,k_\text{B}\,T}-1}\right).
\end{equation} 
At  high temperature, the quantum temperature converges toward
  the  classical  value. When  $T$  decreases  to zero,  $T^\text{C}$
converges  to  the  zero-point  energy  of the  system  (expressed  in
Kelvins).

\begin{figure}[!]
\includegraphics[scale=0.4]{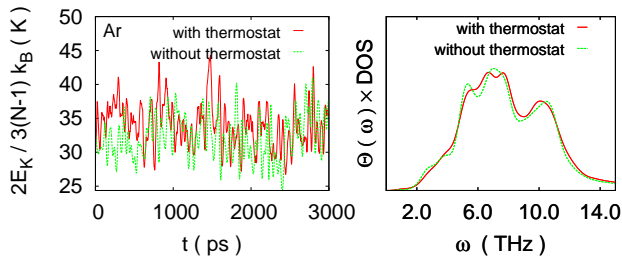}
\caption{ NEMD simulation with $T_\text{1}=T_\text{2}=20$ K for a 5120
  atoms super-cell. Left panel: Twice the kinetic energy per degree of
  freedom  as a  function  of time.  Right  panel: vibrational  phonon
  spectrum times DOS [Eq. (\ref{eq:omega})], obtained from the Fourier
  transform of the  velocity auto-correlation function (VAF). Averages
  were performed over 64 atoms coupled or not to a thermostat. }
\label{fig:nemd}
\end{figure}

In  NEMD, part  of the  system is  not thermostated,  and will  not be
directly coupled to a quantum thermal bath. It is not straightforward if
the interaction  between thermalized  and  non-thermalized parts
will transfer the frequency-dependent energy.  In order to explore the
evolution of  the system  under such conditions,  we performed  a test
simulation  with the  same configuration  as NEMD,  but  coupling both
thermostated  slabs to  quantum thermostats  at the  same temperature.
Figure \ref{fig:nemd}  shows the instantaneous kinetic  energy and the
Fourier transform of the VAF, once the system has reached equilibrium.
Averages were performed  over atoms in different regions  of the cell,
either  thermostated or  not.   As can  be  seen, the  thermostat-free
regions (left panel, dashed  line) have reached an average temperature
in agreement  with the one  imposed in the thermostated  regions (full
red line).   The larger fluctuations  in the thermostated  regions are
due to  the fact  that averages are  computed over smaller  numbers of
atoms.   Moreover, the  Fourier transform  of  the VAF  shown in  Fig.
(\ref{fig:nemd}) (right panel) shows that the same frequency-dependent
energy distribution  is obtained  in both  regions, proving  that a
system can be equilibrated with  a mode-dependent energy by applying a
quantum  thermostat only  to a  subset of  the system.   It  should be
pointed out, however,  that the thermostated regions must  have a size
comparable or greater  than the one of the  ''free'' regions.  If this
is  not the  case,  the  thermostated regions  are  not sufficient  to
thermostat the  free part,  which tends to  relax towards  a classical
energy distribution.  This is in  contrast with the classical case, in
which the thermostating of a few degrees of freedom is, in principle,
sufficient to impose the temperature in an arbitrarily large system.

\section{Thermal conductivity}
\label{therm_cond}
The  methodologies introduced  in Sec.~\ref{methods}  were  applied to
calculate the  thermal conductivity of  solid argon. This  system is
well  documented in the  literature and  can be  modeled with  a simple
Lennard-Jones      interatomic     potential,      with     parameters
$\epsilon/k_\text{B}=120$ K  and $\sigma= 3.4 $  \AA.  All simulations
were carried out with  the \troc \, package.~\cite{troc03} Super-cells
of 256, 1280, 5120 and 10240 atoms were used, and time-steps of 1 or 5
fs.  The potential cutoff was fixed to 4$\sigma$.

\begin{figure}
\includegraphics[scale=0.4]{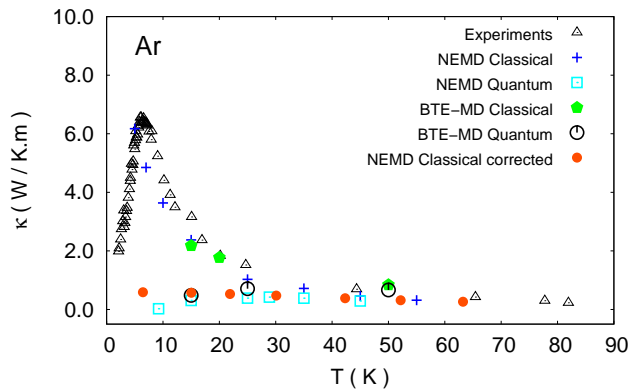}
\caption{Thermal conductivity of solid argon as a function of temperature. Experimental data reported by Christen and Pollack.~\cite{christen_PRB_75}}
\label{fig:argon}
\end{figure}

A  comparison  of  our  results  to  experimental  data  is  shown  in
Fig.   \ref{fig:argon}.   Simulations were  performed  using either  a
classical or a quantum thermostat.   NEMD was implemented with the two
methodologies   mentioned  above,  imposing   either  a   gradient  of
temperature  or  a  flux of  energy.  The  two  methods were  in  full
agreement and are shown here with the same symbols.

NEMD calculations  with the quantum thermostat could  not be performed
below  10  K  because  of  the  difficulty  to  impose  or  measure  a
temperature  gradient  in  this   temperature  range.   To  measure  a
temperature gradient,  we first compute the profile  of kinetic energy
across  the sample,  from which  we deduce  the  classical temperature
$T^\text{C}$,  which serves  to map  the  real temperature  $T$ by  inverting
Eq.   (\ref{eq:temperatures}).  At   low  temperatures   however,  the
classical temperature  converges to the zero-point  energy and becomes
almost  temperature  independent.    Temperature  gradients  are  then
difficult  to  estimate,  requiring  better  statistics,  i.e.  larger
simulation  cells  and  longer  simulation times,  which  limited  our
calculations to temperatures above about 10 K.

We can  see from Fig.  \ref{fig:argon} that all  approaches, classical
and  quantum,  are  in  good  agreement  with  one  another  and  with
experimental data at high temperatures, typically above 40 K.

At  lower  temperatures,  conductivities  computed  with  a  classical
thermostat    remain    in    good   agreement    with    experimental
data~\cite{christen_PRB_75}  down to  about 10  K, while  the computed
quantum conductivities  are much lower. Fitting  the experimental data
to $T^{-n}$, for  temperatures higher than 10 K,  we find $n=1.23$. Our
classical  data present,  in the  same temperature  range,  a slightly
stronger   dependence  with   $n\approx1.32$.  An   agreement  between
classical calculations and experimental data has been obtained as well
by other  authors,~\cite{kaburaki_JAP_07} but is  very surprising since
quantum  effects  on the  specific  heat, which  enters
directly  in   the  expression   of  the  thermal   conductivity  [see
Eq.   (\ref{eq:BT})],   start   at    about   40   K,   as   seen   in
Fig.~\ref{fig:cv_ek}.

The underestimation  of the conductivity using  the quantum thermostat
is   not  due   to  an   inability  of   the  thermostat   to  address
non-equilibrium conditions, since equivalent results are obtained with
the BTE-MD,  which is  an equilibrium based  approach. In  NEMD, aside
from  the  phonon-phonon  scattering  present in  real  materials,  an
additional  phonon-boundary scattering  is present  at  the boundaries
between hot  and cold  sections (if the  system in not  large enough).
The phonon mean free path  is  then reduced, i.e. the phonon lifetimes,
and a  lower thermal conductivity  is obtained.  At  high temperatures
such effect  is less important, as  the mean free path  is governed by
the  phonon-phonon scattering.  The  phonon population  increases with
temperature, increasing the  phonon-phonon scattering, as more phonons
are  present  to   do  the  scattering.~\cite{ashcroft}  However,  the
comparison with  BTE-MD results  suggests that boundary  scattering is
not the main effect that  causes the reduction in thermal conductivity
when using the quantum thermostat.  Indeed, size effects in BTE-MD are
much less important, in the sense that mean free paths are not limited
by the boundaries  of the system. In this case,  a system large enough
must be just  considered in order to ensure  that all modes accessible
to the system are well  described in the simulation.  Our simulations,
have  been performed  for  different  cell sizes  in  order to  ensure
convergence.

\begin{figure}
\includegraphics[scale=0.4]{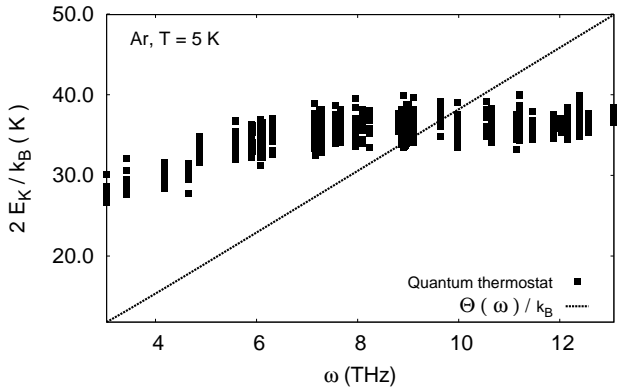}
\caption{Scaled average kinetic energy per  mode obtained from MD simulations compared to the correct quantum harmonic distribution  given by Eq. (\ref{eq:omega}). Crystal argon at 5 K, using a supercell of 256 atoms. The length of the MD simulation was 4 ns.}
\label{fig:leakage_argon}
\end{figure}

\begin{figure}
\includegraphics[scale=0.4]{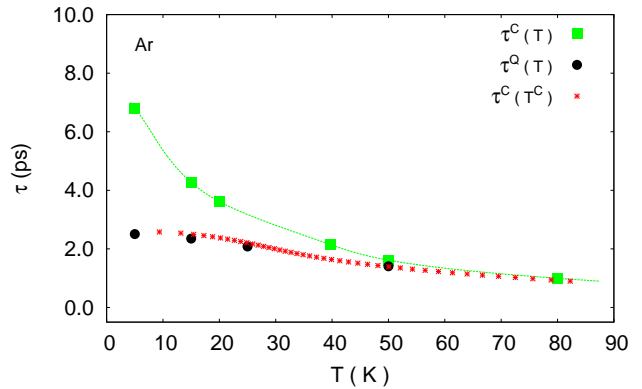}
\caption{Average phonon lifetimes as a function of temperature obtained from BTE-LD calculations.}
\label{fig:life}
\end{figure}

The inability  of the quantum thermostat calculations to describe correctly
the thermal  conductivity is  more probably a  consequence of  the zero
point energy leakage mentioned  in Sec.~\ref{sec:zpe}. The main effect
of this leakage is that the energy is distributed almost homogeneously
among    the    modes,   as    in    the    classical   regime,    as seen in 
Fig.~\ref{fig:leakage_argon}. The system  with the quantum thermostat,
hence,  behaves almost  like  a  classical  system,  but  at  a  higher
temperature. To illustrate this point, we show
in Fig.~\ref{fig:life} the evolution of the average phonon lifetime as
a  function of temperature,  obtained with  the classical  and quantum
thermostats.  Phonon lifetimes obtained  with the  quantum thermostat,
$\tau^\text{Q}$,  are  much  shorter  than  the  classical  lifetimes,
$\tau^\text{C}$,  and the former  can be  obtained from  the latter,
  by   replacing  the   classical  temperature  $T^C$   by  its
corresponding quantum (real) temperature, $T$ , i.e., we have:
\begin{equation}
\label{eq:lifetime}
\langle \tau^\text{Q} (T) \rangle \sim \langle \tau^\text{C} (T^\text{C}) \rangle,
\end{equation}
where      $T$      and      $T^\text{C}$     are      related      by
Eq.  \ref{eq:temperatures}.  This correction corresponds  to the usual
rescaling     of     temperatures     used     for     instance     in
Refs.~\onlinecite{Li1998}  and ~\onlinecite{Jund1999}.   The agreement
between the  corrected classical  lifetimes and the  quantum lifetimes
shown in  Fig.~\ref{fig:life} confirms that the  system described with
the quantum thermostat  is equivalent to a classical  system at higher
temperature.

Using this insight, the thermal conductivity obtained with the quantum
thermostat can  be predicted from the  classical conductivity. Indeed,
if  we assume  that the  specific  heat, group  velocities and  phonon
lifetimes    are    independent,     as    is    often    done    (see
Refs.~\onlinecite{Li1998}, \onlinecite{Jund1999}),      we      can      approximate
Eq. (\ref{eq:BT}) as:
\begin{equation}
\kappa \propto \langle C_\text{ph} \rangle \langle \vec{v}_\text{g}^{\,2} \rangle \langle \tau \rangle.
\label{eq:kappa_average}
\end{equation}

We have  seen in Sec.~\ref{sec:QT-NEMD} that the  quantum thermostat
allows to  reproduce the average specific  heat, so if  we assume that
the group velocity is not strongly affected by quantum effects, we can
write:
\begin{eqnarray}
\label{eq:correct}
\kappa^{\,\text{Q}}\large( T \large) &=& \kappa^{\,\text{C}}\large( T^\text{C} \large)\,\frac{C_{\,\text{v}}^{\,\text{Q}}\large( T \large)}{C_{\,\text{v}}^{\,\text{C}}\large( T^{\,\text{C}} \large)}\,\frac{\langle \tau^{\,\text{Q}}\large( T \large) \rangle}{\langle \tau^{\,\text{C}} \large( T^{\,\text{C}} \large) \rangle}\\
&=& \kappa^{\,\text{C}} \large(T^{\,\text{C}}\large)\, \frac{C_{\,\text{v}}^{\,\text{Q}}\large( T \large)}{3\,(N-1)\,k_\text{B}},\nonumber
\end{eqnarray}
using Eq.  (\ref{eq:lifetime}). The  result of this rescaling is shown
in Fig.  \ref{fig:argon}, as red  full circles.  It is seen that this procedure
 closely  matches  the  conductivities  obtained  with  the
quantum    thermostat.    The    correction    considered    in    Eq.
(\ref{eq:correct}),   even    if   it   is   widely    used   in   the
literature,~\cite{Wang1990,Lee1991,Li1998,Jund1999}  is  known  to  be
oversimplified  and  inaccurate compared  to  the  results  of a  full
quantum approach.~\cite{Turney2009a} On the  other hand, we have shown
here  that   this  correction  fully  explains  the   results  of  the
semi-classical Langevin quantum thermostat.

One  surprising  result  remains  concerning the apparent   absence  of  quantum
effects on the thermal conductivity  of argon. Some authors argue that
quantum effects are not relevant  for argon, even at temperatures well
below the Debye  value, and avoid any correction.~\cite{McGaughey2004}
This simplification is  based on the accuracy of  the classical theory
to  describe properties such as  the nearest  neighbour distance,  the bulk
modulus, and the  cohesive energy of solid noble  gases. The effect of
neglecting  the  zero point  motion  for these  properties, is  less
important in solid argon than in lighter systems.~\cite{ashcroft} But,
from Fig.  \ref{fig:cv_ek},  we know that the heat  capacity starts to
decrease at  about 40 K.   From Eq.  (\ref{eq:kappa_average}),  we see
that an  absence of quantum effects  implies that the  decrease in the
specific  heat  is  compensated  by  an  increase of  the  phonon
lifetime.  However,  an  exact,  or  near exact  compensation  is  not
expected  a  priory and  seems  to be  specific  to  argon, since  for
instance in Si, classical calculations yield conductivities higher than
experimental data.~\cite{Howell2012}

\section{Conclusions}
\label{Conclu}
In this work, various methods based on classical or semi classical molecular dynamics were used to obtain  the  thermal  conductivity of
a very simple system, solid argon. The results of classical nonequilibrium molecular dynamics, of NEMD using a quantum heat bath, and of the Boltzmann transport equation with lifetimes obtained from molecular dynamics were considered, and compared to experimental data.  Very surprisingly, the only method that leads to results in good agreement with  experimental data at low temperature is classical molecular dynamics. It must however been admitted that there are good reasons to believe this agreement in the case of Argon is partially fortuitous, and results from a cancellation of errors between heat capacity and phonon mean free path. Indeed, when an empirical description of the heat capacity is introduced in the Boltzmann transport equation, the agreement with experiments worsens. Moreover, other studies in systems such as diamond silicon have shown that the classical MD results can actually strongly overestimate the thermal conductivity at temperatures below the Debye temperature.

The quantum heat bath method, which was originally thought to be promising, as it assigns the correct zero point energy to the phonon modes, leads to a quite poor agreement with experiments, with a strong underestimation of the thermal conductivity. The basic reason for this discrepancy, which appears both in a BTE approach and in a direct nonequilibrium calculation, is a too short lifetime for the vibrational modes. In turn, the latter can be attributed to the zero point energy leakage issue.  The vibrational amplitude associated with the  zero point energy motion can be exchanged between modes, and can contribute to phonon scattering, which does not correspond to the physical situation in a real quantum system.

A natural question that arises as a result of this work is the existence of a reliable simulation method for computing thermal conductivities in solids below the Debye temperature.  Such a method should be able, if one considers the usual formula of Eq. \ref{eq:BT}, to predict correctly normal mode heat capacities and lifetimes. At present, it appears that no method based on molecular dynamics has the ability to achieve both tasks; classical MD fails on both aspects, while the use of a quantum thermostat results in a strong underestimation of the lifetimes. Ad hoc rescaling of temperatures, or the use of classical phonon lifetimes within a BTE scheme, do not offer any guarantee in terms of reliability or accuracy, although they may work reasonably well for specific systems.  For simple crystal systems, a satisfactory alternative is the use of lattice dynamics techniques for computing the phonon lifetimes, based on quantum perturbation theory and using the cubic term in the expansion of the potential energy.~\cite{debernardi_PRB98,srivastava} Such a method is, however, computationally intensive and tedious. More importantly, it does not seem to be applicable to disordered systems, or even to crystals with complex unit cells. Therefore the calculation of heat conductivity from numerical simulations in such systems at low temperature remains an open challenge.

\begin{acknowledgments}
This work has been supported by the Nanoscience Foundation of Grenoble-France. 
\end{acknowledgments}

\bibliography{bibliography}

\begin{thebibliography}{34}
\expandafter\ifx\csname natexlab\endcsname\relax\def\natexlab#1{#1}\fi
\expandafter\ifx\csname bibnamefont\endcsname\relax
  \def\bibnamefont#1{#1}\fi
\expandafter\ifx\csname bibfnamefont\endcsname\relax
  \def\bibfnamefont#1{#1}\fi
\expandafter\ifx\csname citenamefont\endcsname\relax
  \def\citenamefont#1{#1}\fi
\expandafter\ifx\csname url\endcsname\relax
  \def\url#1{\texttt{#1}}\fi
\expandafter\ifx\csname urlprefix\endcsname\relax\def\urlprefix{URL }\fi
\providecommand{\bibinfo}[2]{#2}
\providecommand{\eprint}[2][]{\url{#2}}

\bibitem[{\citenamefont{Stackhouse and Stixrude}(2010)}]{RMG2010_Stackhouse}
\bibinfo{author}{\bibfnamefont{S.}~\bibnamefont{Stackhouse}} \bibnamefont{and}
  \bibinfo{author}{\bibfnamefont{L.}~\bibnamefont{Stixrude}},
  \bibinfo{journal}{Rev. Mineral. Geochem.} \textbf{\bibinfo{volume}{71}},
  \bibinfo{pages}{253} (\bibinfo{year}{2010}).

\bibitem[{\citenamefont{Jund and Jullien}(1999{\natexlab{a}})}]{PRB1999_Jund}
\bibinfo{author}{\bibfnamefont{P.}~\bibnamefont{Jund}} \bibnamefont{and}
  \bibinfo{author}{\bibfnamefont{R.}~\bibnamefont{Jullien}},
  \bibinfo{journal}{Phys. Rev. B} \textbf{\bibinfo{volume}{59}},
  \bibinfo{pages}{13707} (\bibinfo{year}{1999}{\natexlab{a}}).

\bibitem[{\citenamefont{M\"{u}ller-Plathe}(1997)}]{JCP1997_Plathe}
\bibinfo{author}{\bibfnamefont{F.}~\bibnamefont{M\"{u}ller-Plathe}},
  \bibinfo{journal}{J. Chem. Phys.} \textbf{\bibinfo{volume}{106}},
  \bibinfo{pages}{6082} (\bibinfo{year}{1997}).

\bibitem[{\citenamefont{Turney et~al.}(2009{\natexlab{a}})\citenamefont{Turney,
  McGaughey, and Amon}}]{Turney2009a}
\bibinfo{author}{\bibfnamefont{J.}~\bibnamefont{Turney}},
  \bibinfo{author}{\bibfnamefont{A.}~\bibnamefont{McGaughey}},
  \bibnamefont{and} \bibinfo{author}{\bibfnamefont{C.}~\bibnamefont{Amon}},
  \bibinfo{journal}{Phys. Rev. B} \textbf{\bibinfo{volume}{79}},
  \bibinfo{pages}{224305} (\bibinfo{year}{2009}{\natexlab{a}}).

\bibitem[{\citenamefont{Li et~al.}(1998)\citenamefont{Li, Porter, and
  Yip}}]{Li1998}
\bibinfo{author}{\bibfnamefont{J.}~\bibnamefont{Li}},
  \bibinfo{author}{\bibfnamefont{L.}~\bibnamefont{Porter}}, \bibnamefont{and}
  \bibinfo{author}{\bibfnamefont{S.}~\bibnamefont{Yip}}, \bibinfo{journal}{J.
  Nucl. Mater.} \textbf{\bibinfo{volume}{255}}, \bibinfo{pages}{139}
  (\bibinfo{year}{1998}).

\bibitem[{\citenamefont{Jund and Jullien}(1999{\natexlab{b}})}]{Jund1999}
\bibinfo{author}{\bibfnamefont{P.}~\bibnamefont{Jund}} \bibnamefont{and}
  \bibinfo{author}{\bibfnamefont{R.}~\bibnamefont{Jullien}},
  \bibinfo{journal}{Phys. Rev. B} \textbf{\bibinfo{volume}{59}},
  \bibinfo{pages}{13707} (\bibinfo{year}{1999}{\natexlab{b}}).

\bibitem[{\citenamefont{Kaburaki et~al.}(2007)\citenamefont{Kaburaki, Li, Yip,
  and Kimizuka}}]{kaburaki_JAP_07}
\bibinfo{author}{\bibfnamefont{H.}~\bibnamefont{Kaburaki}},
  \bibinfo{author}{\bibfnamefont{J.}~\bibnamefont{Li}},
  \bibinfo{author}{\bibfnamefont{S.}~\bibnamefont{Yip}}, \bibnamefont{and}
  \bibinfo{author}{\bibfnamefont{H.}~\bibnamefont{Kimizuka}},
  \bibinfo{journal}{J. Appl. Phys.} \textbf{\bibinfo{volume}{102}},
  \bibinfo{pages}{6} (\bibinfo{year}{2007}).

\bibitem[{\citenamefont{McGaughey and
  Kaviany}(2004{\natexlab{a}})}]{McGaughey2004}
\bibinfo{author}{\bibfnamefont{A.}~\bibnamefont{McGaughey}} \bibnamefont{and}
  \bibinfo{author}{\bibfnamefont{M.}~\bibnamefont{Kaviany}},
  \bibinfo{journal}{Int. J. Heat Mass Transfer} \textbf{\bibinfo{volume}{47}},
  \bibinfo{pages}{1783} (\bibinfo{year}{2004}{\natexlab{a}}).

\bibitem[{\citenamefont{Srivastava}(1990)}]{srivastava}
\bibinfo{author}{\bibfnamefont{G.~P.} \bibnamefont{Srivastava}},
  \emph{\bibinfo{title}{The Physics of phonons}} (\bibinfo{publisher}{Taylor \&
  Francis Group}, \bibinfo{year}{1990}).

\bibitem[{\citenamefont{Dammak et~al.}(2009)\citenamefont{Dammak, Chalopin,
  Laroche, Hayoun, and Greffet}}]{PRL2009_Dammak}
\bibinfo{author}{\bibfnamefont{H.}~\bibnamefont{Dammak}},
  \bibinfo{author}{\bibfnamefont{Y.}~\bibnamefont{Chalopin}},
  \bibinfo{author}{\bibfnamefont{M.}~\bibnamefont{Laroche}},
  \bibinfo{author}{\bibfnamefont{M.}~\bibnamefont{Hayoun}}, \bibnamefont{and}
  \bibinfo{author}{\bibfnamefont{J.-J.} \bibnamefont{Greffet}},
  \bibinfo{journal}{Phys. Rev. Lett.} \textbf{\bibinfo{volume}{103}},
  \bibinfo{pages}{190601} (\bibinfo{year}{2009}).

\bibitem[{\citenamefont{Ceriotti et~al.}(2009)\citenamefont{Ceriotti, Bussi,
  and Parrinello}}]{PRL2009_Parrinello}
\bibinfo{author}{\bibfnamefont{M.}~\bibnamefont{Ceriotti}},
  \bibinfo{author}{\bibfnamefont{G.}~\bibnamefont{Bussi}}, \bibnamefont{and}
  \bibinfo{author}{\bibfnamefont{M.}~\bibnamefont{Parrinello}},
  \bibinfo{journal}{Phys. Rev. Lett.} \textbf{\bibinfo{volume}{103}},
  \bibinfo{pages}{030603} (\bibinfo{year}{2009}).

\bibitem[{\citenamefont{Barrat and Rodney}(2011)}]{JSP2011_Barrat_Rodney}
\bibinfo{author}{\bibfnamefont{J.-L.} \bibnamefont{Barrat}} \bibnamefont{and}
  \bibinfo{author}{\bibfnamefont{D.}~\bibnamefont{Rodney}},
  \bibinfo{journal}{J. Stat. Phys.} \textbf{\bibinfo{volume}{144}},
  \bibinfo{pages}{679} (\bibinfo{year}{2011}).

\bibitem[{\citenamefont{Savin et~al.}(2012)\citenamefont{Savin, Kosevich, and
  Cantarero}}]{Savin2012}
\bibinfo{author}{\bibfnamefont{A.}~\bibnamefont{Savin}},
  \bibinfo{author}{\bibfnamefont{Y.}~\bibnamefont{Kosevich}}, \bibnamefont{and}
  \bibinfo{author}{\bibfnamefont{A.}~\bibnamefont{Cantarero}},
  \bibinfo{journal}{Phys. Rev. B} \textbf{\bibinfo{volume}{86}},
  \bibinfo{pages}{064305} (\bibinfo{year}{2012}).

\bibitem[{\citenamefont{Turney et~al.}(2009{\natexlab{b}})\citenamefont{Turney,
  Landry, McGaughey, and Amon}}]{PRB09_turney}
\bibinfo{author}{\bibfnamefont{J.~E.} \bibnamefont{Turney}},
  \bibinfo{author}{\bibfnamefont{E.~S.} \bibnamefont{Landry}},
  \bibinfo{author}{\bibfnamefont{A.~J.~H.} \bibnamefont{McGaughey}},
  \bibnamefont{and} \bibinfo{author}{\bibfnamefont{C.~H.} \bibnamefont{Amon}},
  \bibinfo{journal}{Phys. Rev. B} \textbf{\bibinfo{volume}{79}},
  \bibinfo{pages}{064301} (\bibinfo{year}{2009}{\natexlab{b}}).

\bibitem[{\citenamefont{McGaughey and
  Kaviany}(2004{\natexlab{b}})}]{McGaughey2004b}
\bibinfo{author}{\bibfnamefont{A.}~\bibnamefont{McGaughey}} \bibnamefont{and}
  \bibinfo{author}{\bibfnamefont{M.}~\bibnamefont{Kaviany}},
  \bibinfo{journal}{Phys. Rev. B} \textbf{\bibinfo{volume}{69}},
  \bibinfo{pages}{094303} (\bibinfo{year}{2004}{\natexlab{b}}), ISSN
  \bibinfo{issn}{1098-0121}.

\bibitem[{\citenamefont{Green}(1954)}]{Green1954}
\bibinfo{author}{\bibfnamefont{M.~S.} \bibnamefont{Green}},
  \bibinfo{journal}{J. Chem. Phys.} \textbf{\bibinfo{volume}{22}},
  \bibinfo{pages}{398} (\bibinfo{year}{1954}).

\bibitem[{\citenamefont{Kubo}(1957)}]{Kubo1957}
\bibinfo{author}{\bibfnamefont{R.}~\bibnamefont{Kubo}}, \bibinfo{journal}{J.
  Phys. Soc. Jpn.} \textbf{\bibinfo{volume}{12}}, \bibinfo{pages}{570}
  (\bibinfo{year}{1957}).

\bibitem[{\citenamefont{Risken}(1989)}]{risken}
\bibinfo{author}{\bibfnamefont{H.}~\bibnamefont{Risken}},
  \emph{\bibinfo{title}{The Fokker-Planck Equation}}
  (\bibinfo{publisher}{Springer-Verlag}, \bibinfo{year}{1989}).

\bibitem[{\citenamefont{Grest et~al.}(1986)\citenamefont{Grest, Kremer, and
  Carlo}}]{Grest1986}
\bibinfo{author}{\bibfnamefont{G.~S.} \bibnamefont{Grest}},
  \bibinfo{author}{\bibfnamefont{K.}~\bibnamefont{Kremer}}, \bibnamefont{and}
  \bibinfo{author}{\bibfnamefont{M.}~\bibnamefont{Carlo}},
  \bibinfo{journal}{Phys. Rev. A} \textbf{\bibinfo{volume}{33}},
  \bibinfo{pages}{3628} (\bibinfo{year}{1986}).

\bibitem[{\citenamefont{Oppenheim and Schafer}(1999)}]{Oppenheim2009}
\bibinfo{author}{\bibfnamefont{A.~V.} \bibnamefont{Oppenheim}}
  \bibnamefont{and} \bibinfo{author}{\bibfnamefont{R.~W.}
  \bibnamefont{Schafer}}, \emph{\bibinfo{title}{Discrete-Time Signal Processing
  (2nd Edition)}} (\bibinfo{publisher}{Prentice Hall, Englewoods Cliffs, NJ},
  \bibinfo{year}{1999}).

\bibitem[{\citenamefont{Miller et~al.}(1989)\citenamefont{Miller, Hase, and
  Darling}}]{Miller1989}
\bibinfo{author}{\bibfnamefont{W.~H.} \bibnamefont{Miller}},
  \bibinfo{author}{\bibfnamefont{W.~L.} \bibnamefont{Hase}}, \bibnamefont{and}
  \bibinfo{author}{\bibfnamefont{C.~L.} \bibnamefont{Darling}},
  \bibinfo{journal}{J. Chem. Phys.} \textbf{\bibinfo{volume}{91}},
  \bibinfo{pages}{2863} (\bibinfo{year}{1989}).

\bibitem[{\citenamefont{Alimi et~al.}(1992)\citenamefont{Alimi, García-Vela,
  and Gerber}}]{Alimi1992}
\bibinfo{author}{\bibfnamefont{R.}~\bibnamefont{Alimi}},
  \bibinfo{author}{\bibfnamefont{A.}~\bibnamefont{García-Vela}},
  \bibnamefont{and} \bibinfo{author}{\bibfnamefont{R.~B.}
  \bibnamefont{Gerber}}, \bibinfo{journal}{J. Chem. Phys.}
  \textbf{\bibinfo{volume}{96}}, \bibinfo{pages}{2034} (\bibinfo{year}{1992}).

\bibitem[{\citenamefont{Ben-Nun and Levine}(1996)}]{Ben-Nun1996}
\bibinfo{author}{\bibfnamefont{M.}~\bibnamefont{Ben-Nun}} \bibnamefont{and}
  \bibinfo{author}{\bibfnamefont{R.~D.} \bibnamefont{Levine}},
  \bibinfo{journal}{J. Chem. Phys.} \textbf{\bibinfo{volume}{105}},
  \bibinfo{pages}{8136} (\bibinfo{year}{1996}).

\bibitem[{\citenamefont{Habershon and Manolopoulos}(2009)}]{Habershon2009}
\bibinfo{author}{\bibfnamefont{S.}~\bibnamefont{Habershon}} \bibnamefont{and}
  \bibinfo{author}{\bibfnamefont{D.~E.} \bibnamefont{Manolopoulos}},
  \bibinfo{journal}{J. Chem. Phys.} \textbf{\bibinfo{volume}{131}},
  \bibinfo{pages}{244518} (\bibinfo{year}{2009}).

\bibitem[{\citenamefont{Czak\'{o} et~al.}(2010)\citenamefont{Czak\'{o},
  Kaledin, and Bowman}}]{Czako2010}
\bibinfo{author}{\bibfnamefont{G.}~\bibnamefont{Czak\'{o}}},
  \bibinfo{author}{\bibfnamefont{A.~L.} \bibnamefont{Kaledin}},
  \bibnamefont{and} \bibinfo{author}{\bibfnamefont{J.~M.}
  \bibnamefont{Bowman}}, \bibinfo{journal}{J. Chem. Phys.}
  \textbf{\bibinfo{volume}{132}}, \bibinfo{pages}{164103}
  (\bibinfo{year}{2010}).

\bibitem[{\citenamefont{Ceriotti et~al.}(2010)\citenamefont{Ceriotti, Miceli,
  Pietropaolo, Colognesi, Nale, Catti, Bernasconi, and
  Parrinello}}]{PRB2010_Parrinello_H}
\bibinfo{author}{\bibfnamefont{M.}~\bibnamefont{Ceriotti}},
  \bibinfo{author}{\bibfnamefont{G.}~\bibnamefont{Miceli}},
  \bibinfo{author}{\bibfnamefont{A.}~\bibnamefont{Pietropaolo}},
  \bibinfo{author}{\bibfnamefont{D.}~\bibnamefont{Colognesi}},
  \bibinfo{author}{\bibfnamefont{A.}~\bibnamefont{Nale}},
  \bibinfo{author}{\bibfnamefont{M.}~\bibnamefont{Catti}},
  \bibinfo{author}{\bibfnamefont{M.}~\bibnamefont{Bernasconi}},
  \bibnamefont{and}
  \bibinfo{author}{\bibfnamefont{M.}~\bibnamefont{Parrinello}},
  \bibinfo{journal}{Phys. Rev. B} \textbf{\bibinfo{volume}{82}},
  \bibinfo{pages}{174306} (\bibinfo{year}{2010}).

\bibitem[{\citenamefont{Klein et~al.}(1969)\citenamefont{Klein, Horton, and
  Feldman}}]{cv_exp}
\bibinfo{author}{\bibfnamefont{M.~L.} \bibnamefont{Klein}},
  \bibinfo{author}{\bibfnamefont{G.~K.} \bibnamefont{Horton}},
  \bibnamefont{and} \bibinfo{author}{\bibfnamefont{J.~L.}
  \bibnamefont{Feldman}}, \bibinfo{journal}{Phys. Rev.}
  \textbf{\bibinfo{volume}{184}}, \bibinfo{pages}{968} (\bibinfo{year}{1969}).

\bibitem[{\citenamefont{Rurali and Hern\'andez}(2003)}]{troc03}
\bibinfo{author}{\bibfnamefont{R.}~\bibnamefont{Rurali}} \bibnamefont{and}
  \bibinfo{author}{\bibfnamefont{E.}~\bibnamefont{Hern\'andez}},
  \bibinfo{journal}{Comput. Mat. Sci} \textbf{\bibinfo{volume}{28}},
  \bibinfo{pages}{85} (\bibinfo{year}{2003}).

\bibitem[{\citenamefont{Christen and Pollack}(1975)}]{christen_PRB_75}
\bibinfo{author}{\bibfnamefont{D.~K.} \bibnamefont{Christen}} \bibnamefont{and}
  \bibinfo{author}{\bibfnamefont{G.~L.} \bibnamefont{Pollack}},
  \bibinfo{journal}{Phys. Rev. B} \textbf{\bibinfo{volume}{12}},
  \bibinfo{pages}{3380} (\bibinfo{year}{1975}).

\bibitem[{\citenamefont{Ashcroft and Mermin}(1976)}]{ashcroft}
\bibinfo{author}{\bibfnamefont{N.~W.} \bibnamefont{Ashcroft}} \bibnamefont{and}
  \bibinfo{author}{\bibfnamefont{N.~D.} \bibnamefont{Mermin}},
  \emph{\bibinfo{title}{Solid State Physics}} (\bibinfo{publisher}{Thomson
  Learning , Inc}, \bibinfo{year}{1976}).

\bibitem[{\citenamefont{Wang et~al.}(1990)\citenamefont{Wang, Chan, and
  Ho}}]{Wang1990}
\bibinfo{author}{\bibfnamefont{C.}~\bibnamefont{Wang}},
  \bibinfo{author}{\bibfnamefont{C.}~\bibnamefont{Chan}}, \bibnamefont{and}
  \bibinfo{author}{\bibfnamefont{K.}~\bibnamefont{Ho}}, \bibinfo{journal}{Phys.
  Rev. B} \textbf{\bibinfo{volume}{42}}, \bibinfo{pages}{11276}
  (\bibinfo{year}{1990}).

\bibitem[{\citenamefont{Lee et~al.}(1991)\citenamefont{Lee, Biswas, Soukoulis,
  Wang, Chan, and Ho}}]{Lee1991}
\bibinfo{author}{\bibfnamefont{Y.}~\bibnamefont{Lee}},
  \bibinfo{author}{\bibfnamefont{R.}~\bibnamefont{Biswas}},
  \bibinfo{author}{\bibfnamefont{C.}~\bibnamefont{Soukoulis}},
  \bibinfo{author}{\bibfnamefont{C.}~\bibnamefont{Wang}},
  \bibinfo{author}{\bibfnamefont{C.}~\bibnamefont{Chan}}, \bibnamefont{and}
  \bibinfo{author}{\bibfnamefont{K.}~\bibnamefont{Ho}}, \bibinfo{journal}{Phys.
  Rev. B} \textbf{\bibinfo{volume}{43}}, \bibinfo{pages}{6573}
  (\bibinfo{year}{1991}).

\bibitem[{\citenamefont{Howell}(2012)}]{Howell2012}
\bibinfo{author}{\bibfnamefont{P.~C.} \bibnamefont{Howell}},
  \bibinfo{journal}{J. Chem. Phys.} \textbf{\bibinfo{volume}{137}},
  \bibinfo{pages}{224111} (\bibinfo{year}{2012}).

\bibitem[{\citenamefont{Debernardi}(1998)}]{debernardi_PRB98}
\bibinfo{author}{\bibfnamefont{A.}~\bibnamefont{Debernardi}},
  \bibinfo{journal}{Phys. Rev. B} \textbf{\bibinfo{volume}{57}},
  \bibinfo{pages}{12847} (\bibinfo{year}{1998}).

\end{thebibliography}

\end{document}